\documentstyle[psfig,amstex]{mn}
\def\la{Ly$\alpha$}
\def\etal{et~al.}
\def\ang{\AA}               
\def\spose#1{\hbox to 0pt{#1\hss}}
\def\lta{\mathrel{\spose{\lower 3pt\hbox{$\mathchar"218$}}
     \raise 2.0pt\hbox{$\mathchar"13C$}}}
\def\gta{\mathrel{\spose{\lower 3pt\hbox{$\mathchar"218$}}
     \raise 2.0pt\hbox{$\mathchar"13E$}}}

\def\aips{{\sc aips}}

\title[Jet--cloud interaction in 3C34]{A jet--cloud interaction in 3C34 at
redshift $\mathbf {z = 0.69}$} 

\author[P.~N.~Best, M.~S.~Longair and
H.~J.~A.~R\"ottgering]{P.~N.~Best$^{1,2}$, M.~S.~Longair$^1$ and
H.~J.~A.~R\"ottgering$^2$\\ $^1$ Cavendish Laboratory, Madingley Road,
Cambridge, CB3 0HE, United Kingdom \\ $^2$ Sterrewacht Leiden, Huygens
Laboratory, Postbus 9513, 2300 RA Leiden, The Netherlands}

\pagerange{\pageref{firstpage}--\pageref{lastpage}}
\pubyear{1996}
\begin{document}
\label{firstpage}

\maketitle

\begin{abstract}
\noindent We report the detection of a strong jet--cloud interaction at a
distance of 120~kpc from the nucleus of the radio galaxy 3C34, which has
redshift $z=0.69$. Hubble Space Telescope images of the radio galaxy show
a long narrow region of blue emission orientated along the radio axis and
directed towards a radio hotspot. The William Herschel Telescope has been
used to provide long--slit spectroscopic data of this object, and infrared
observations made with the United Kingdom InfraRed Telescope have enabled
its spectral energy distribution to be modelled. We propose that the
aligned emission is associated with a region of massive star--formation,
induced by the passage of the radio jet through a galaxy within the
cluster surrounding 3C34. A star--formation rate of about 100 $M_{\odot}$
yr$^{-1}$ is required, similar to the values necessary to produce the
alignment effect in high--redshift radio galaxies. The consequences of
this result for models of star formation in distant radio galaxies are
discussed.

\end{abstract}

\section{Introduction}
\label{intro}

In 1985 it was discovered that the optical colours and line ratios of
Minkowski's object, a peculiar galaxy lying along the radio jet of the
source PKS\,0123--016 at redshift $z = 0.019$, are consistent with it
recently having undergone a period of intense star--formation
\cite{bro85,bre85}. This was interpreted as having been triggered by the
interaction of the radio jet with a gas--rich cloud along its
path. Similar phenomena have also been observed in other low redshift
galaxies, most notably in the lobe of the double radio source 3C285 ($z =
0.0794$, van Breugel and Dey 1993)\nocite{bre93}.

In 1987, Chambers \etal\ and McCarthy \etal \nocite{cha87,mcc87}
discovered that the optical emission of powerful radio galaxies at
redshift $z \gta 0.6$ is elongated and aligned along the radio axis. A
natural interpretation of these aligned blue structures was to associate
them with regions of massive star formation induced by shocks associated
with the passage of the radio jet (eg. Rees 1989)\nocite{ree89}. However,
despite various theoretical works suggesting that powerful jets in these
high redshift sources are capable of producing the observed levels of
bright aligned structures \cite{dey89,beg89,dal90}, direct evidence for
the presence of young stars has been scarce. Indeed, the discovery that
the extended optical emission is frequently polarised (eg. Dey and Spinrad
\shortcite{dey96} and references therein) indicates that at least a
proportion of the aligned light must be associated with light scattered
from an obscured active nucleus.

We are undertaking an investigation of an almost complete sample of 28
radio galaxies with redshifts $0.6 < z < 1.8$ from the 3CR catalogue of
Laing \etal\ \shortcite{lai83}. In this paper, we discuss the case of the
radio galaxy 3C34, which is a typical FRII double radio source
\cite{fan74} whose host galaxy is the brightest member of a rich compact
cluster of galaxies \cite{mcc88} at redshift $z = 0.689$. Hubble Space
Telescope observations provide evidence for a jet--cloud interaction
having occurred in this radio source, the cloud in this case being
associated with the interstellar gas of a galaxy within the cluster.

The observations are presented in Section~\ref{observs}, together with
details of the data reduction. In Section~\ref{jetclint}, we discuss the
evidence for a jet--cloud interaction having occurred in this radio
source. We consider the various causes of the observed optical alignment
in Section~\ref{aligns}, and show that the observations are consistent
with jet--induced star formation models. Our conclusions are presented in
Section~\ref{concs}.

\begin{figure*}
\centerline{
\psfig{figure=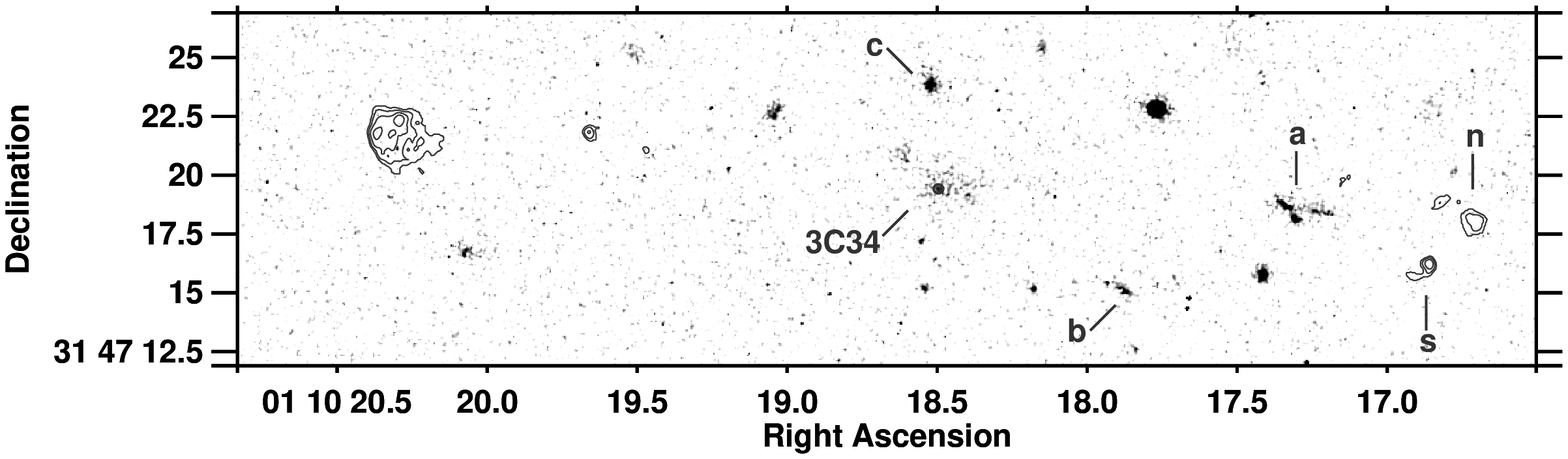,width=15cm,clip=,angle=90}
}
\centerline{
\psfig{figure=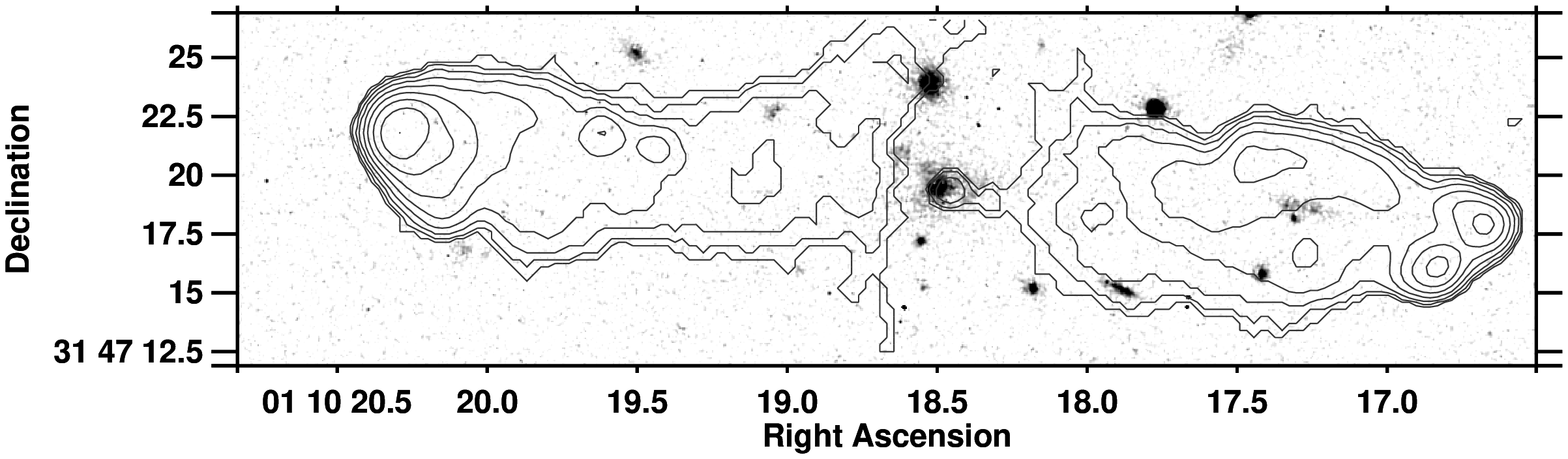,width=15cm,clip=,angle=90}
}
\caption{\label{hstims} (a -- top) An image of 3C34 at 545nm taken using
the Hubble Space Telescope, with contours of the VLA A--array radio
emission at 8.4~GHz overlaid. Contour levels are 120\,$\mu$Jy $\times
(2,4,8,16)$. (b -- bottom) An HST image of 3C34 at 865nm, overlaid with
contours of the radio emission at 4.8~GHz as seen using B and C arrays of
the VLA (Johnson \etal\ 1995). Contour levels are 120\,$\mu$Jy $\times
(1,2,4,8,16,32,64,128)$. All co--ordinates are measured in equinox J2000.}
\end{figure*}

\section{Observations}
\label{observs}

The field of 3C34 was imaged using the Wide--Field Planetary Camera II
(WFPC2) of the Hubble Space Telescope (HST) for 1700 seconds through each
of the two filters f555W and f785LP, centred at wavelengths of 545 and
865\,nm and corresponding to rest--frame near--ultraviolet and visible
wavelengths respectively. The data were reduced according to the standard
Space Telescope Science Institute pipeline \cite{lau89}. Radio data with
comparable angular resolution to the HST images were obtained using the
Very Large Array radio interferometer (VLA) at 8.4~GHz, for 44 minutes in
the A--array configuration and 30 minutes using the C--array. The \aips\
software provided by the National Radio Astronomy Observatory was used to
reduce these data \cite{per89}. In addition, the field was observed in
each of the infrared J ($1.2\mu$m) and K ($2.2\mu$m) wavebands using UKIRT
for 54 minutes, in August 1994. The reader is referred to Best \etal\
\shortcite{bes97c} for a full discussion of the data reduction.

\begin{figure*}
\centerline{
\psfig{figure=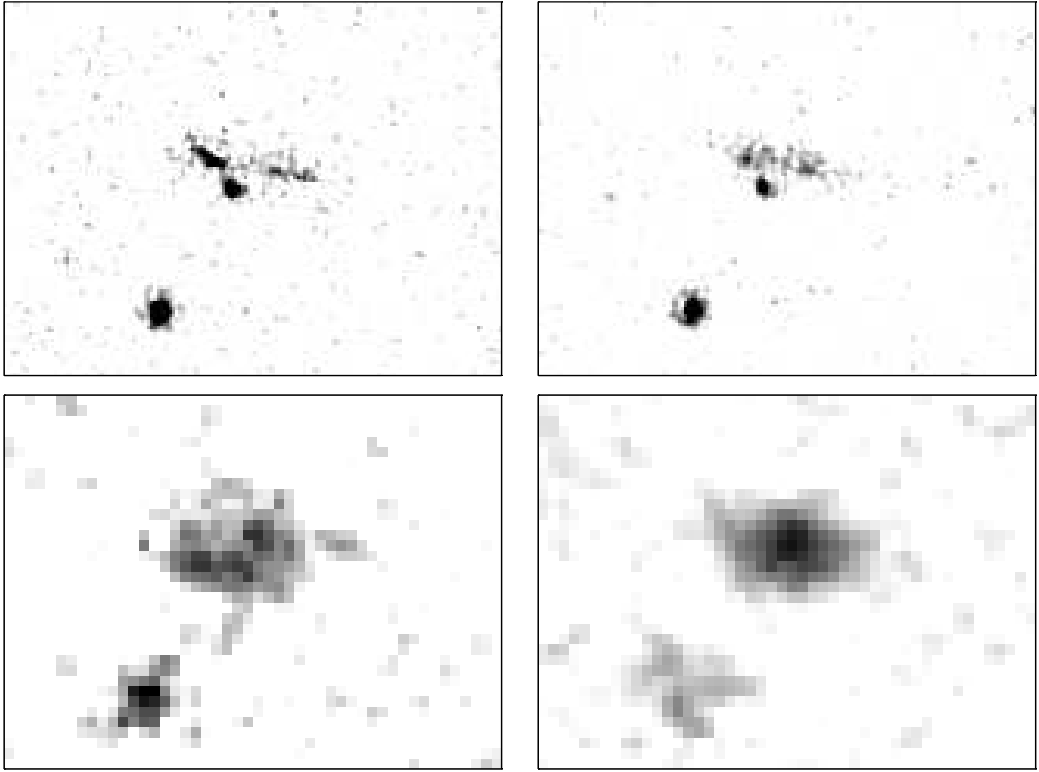,width=13.8cm,clip=,angle=-90}
}
\caption{\label{azoom} Enlarged images of object `a': (a: upper left) HST
image through f555W filter; (b: upper right) HST image through f785LP
filter; (c: lower left) J--band (1.2$\mu$m) UKIRT image; (d: lower right)
K--band (2.2$\mu$m) UKIRT image. All four images are 9.6 by 7.2 arcsec.}
\end{figure*}

\begin{figure*}
\centerline{
\psfig{figure=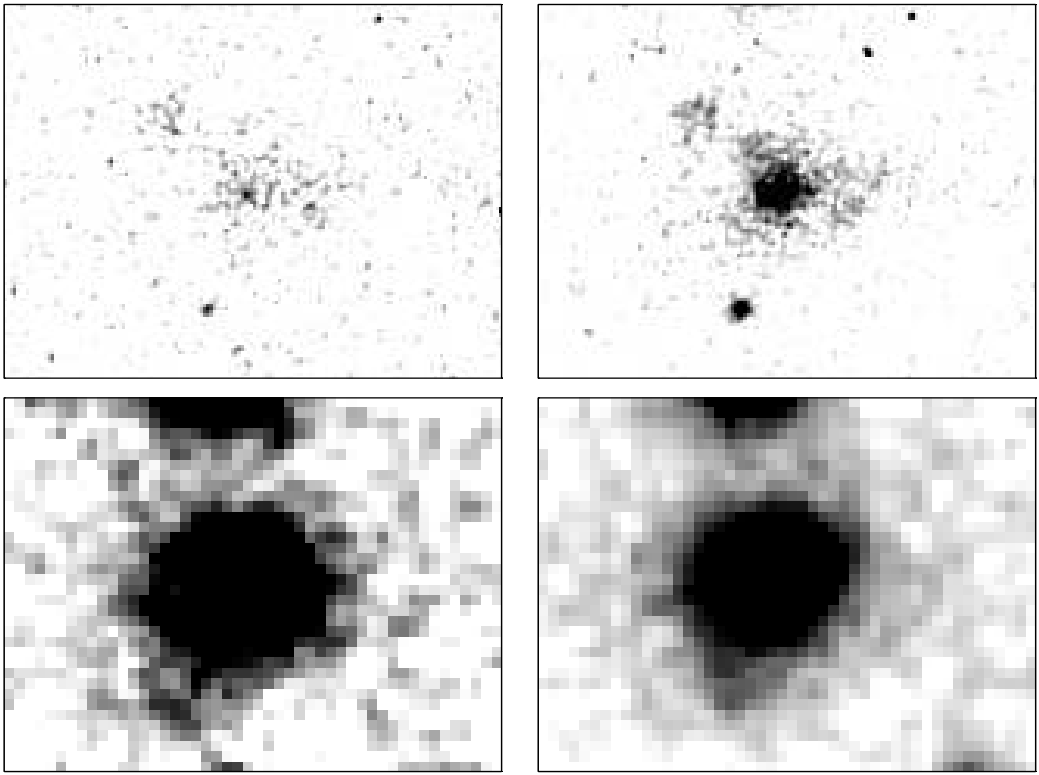,width=13.8cm,clip=,angle=-90}
}
\caption{\label{34zoom} Enlarged images of the galaxy associated with 3C34
in four wavebands: (a: upper left) HST image through f555W filter; (b:
upper right) HST image through f785LP filter; (c: lower left) J--band
(1.2$\mu$m) UKIRT image; (d: lower right) K--band (2.2$\mu$m) UKIRT
image. Both angular sizes and grey-scale levels are the same as in
Figure~\ref{azoom}.}
\end{figure*}

Figure~\ref{hstims}a shows the HST image of the galaxy as observed through
the f555W filter, with the radio contours from the A--array observations
overlaid. In Figure~\ref{hstims}b we present a deep 4.8~GHz radio map
(provided courtesy of Dr J.P.~Leahy) overlaid upon the HST image through
the f785LP filter. Of particular interest is the emission feature,
hereafter object `a', at RA: 01 10 17.3, Dec: 31 47 18 (J2000): two long,
narrow regions of intense blue emission lie directly along a line from the
radio core to the northern of the pair of radio hotspots in the western
lobe (hotspot `n'). A further emission knot lies just to the south of
these. In Figure~\ref{azoom} we present enlarged images of object `a' at
all four wavelengths to show the wavelength--dependent morphology of this
region. For comparison, in Figure~\ref{34zoom} we present enlarged images
of the host radio galaxy 3C34 on the same angular scale, and with the same
grey-scale levels.

Object `a' lies 15 arcsec from the host galaxy 3C34, corresponding to a
projected distance of 120~kpc, assuming $H_0 = 50$\,km s$^{-1}$ Mpc$^{-1}$
and $\Omega = 1$. The large length--to--width extension along the radio
axis of these optical emission regions strongly suggests an association
with the radio source, and that they are produced by an interaction of the
radio jet powering the outer hotspots with a cloud of ambient gas, rather
than by a beam of ionising photons from an obscured quasar nucleus.

In December 1995 a long--slit spectrum of this source was obtained using
the ISIS spectrograph on the William Herschel Telescope (WHT). A slit of
width 2.5 arcsec was orientated at position angle 86.5 degrees, containing
both the host radio galaxy and object `a'. The R158R and R158B gratings
were used in the red and blue arms of the spectrograph respectively, in
conjunction with TEK CCD's. Since only a short (900 second) observation
was made, the read--noise was reduced by binning the data into pixel pairs
in the wavelength direction. The total useful wavelength range of the two
ISIS arms was about 3500 to 8500 \AA, with a small gap from 5700 to 6000
\AA. The effective spectral resolution FWHM was 6 \AA, and the seeing was
about one arcsec. The standard star SP0105+625 was observed immediately
before 3C34 to provide accurate flux calibration, whilst wavelength
calibration was obtained by observations of a Cu--Ar arc lamp. The spectra
were reduced using standard IRAF routines.

\begin{figure}
\centerline{
\psfig{figure=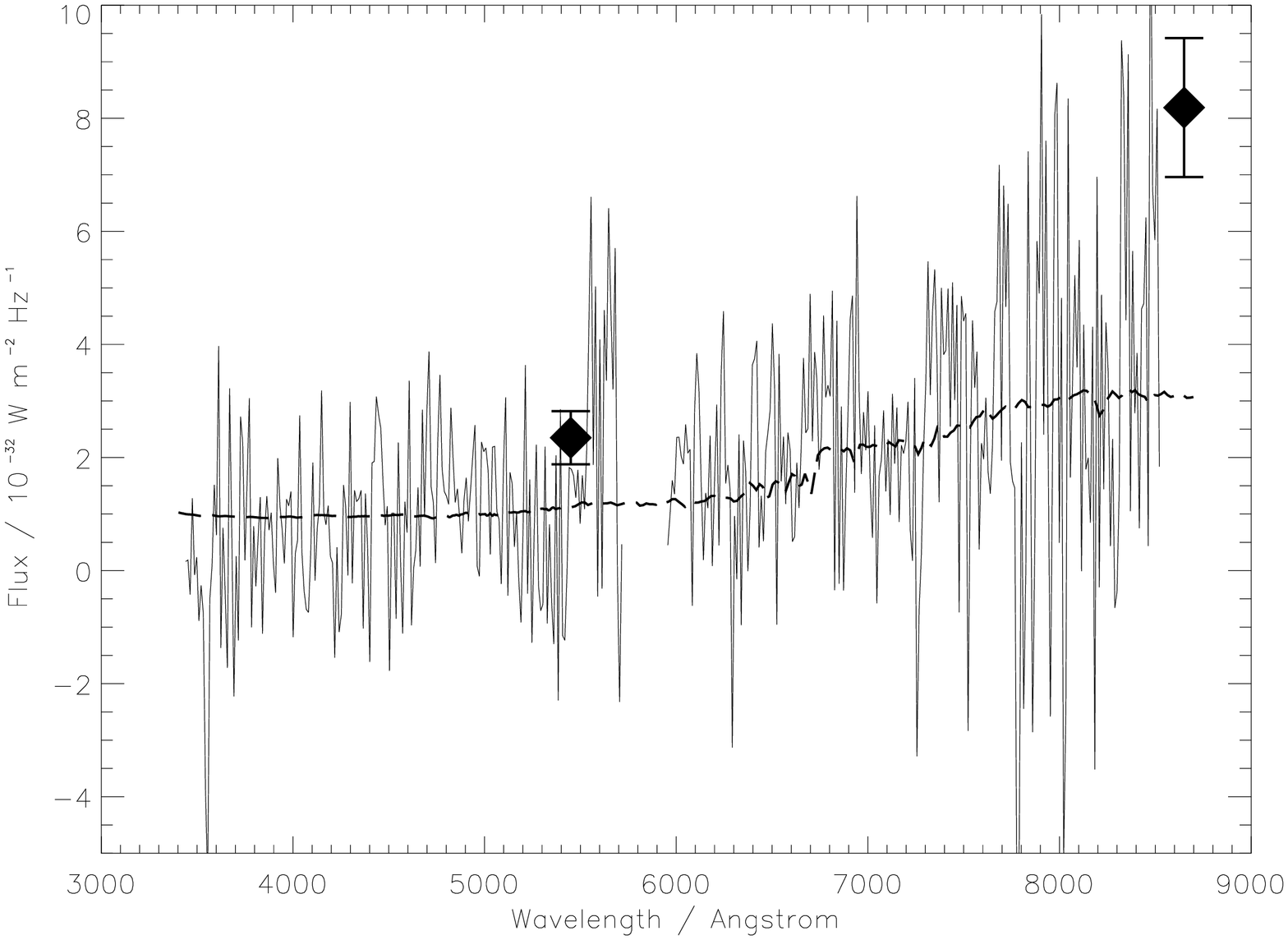,width=7.8cm,clip=,angle=90}
}
\centerline{
\psfig{figure=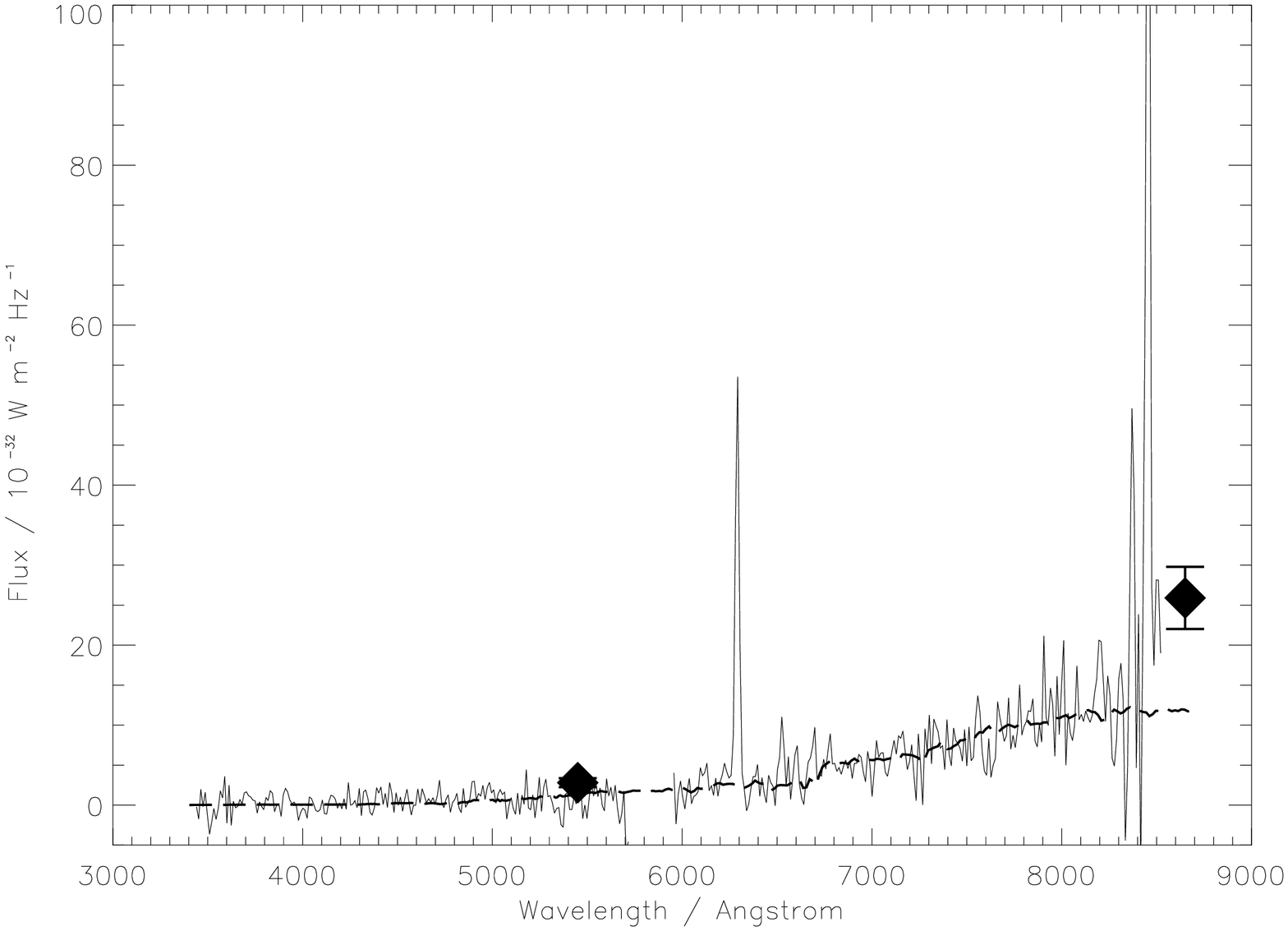,width=7.8cm,clip=,angle=90}
}
\caption{\label{specs}(a -- top) A 900s spectrum of object `a' taken using
the ISIS spectrograph on the WHT. Both red and blue arms are shown. (b --
bottom) A spectrum of 3C34, taken through the same long slit.  In each
case, the broad band HST flux densities are marked by filled diamonds (for
3C34 these have been corrected for line contamination, to give an estimate
of the continuum flux density). These flux densities indicate that the
slit didn't pass through the centre of these objects --- see text for more
details. The dashed lines show the fits to the spectrum using the stellar
synthesis models discussed in Section~\ref{undergal}.}
\end{figure}

The spectrum of object `a' is presented in Figure~\ref{specs}a. It can be
seen that this shows no evidence of any line emission, making it
impossible to derive a redshift for the source. Line emission of both
[OII]~3727 and [OIII]~5007 at a redshift of 0.689 would fall within this
spectrum. For comparison, in Figure~\ref{specs}b we present the spectrum
of the host radio galaxy 3C34, which also lay within the slit. These
strong emission lines dominate this spectrum, with flux densities of:
${\rm f}_{\rm [OII]\,3727} = (9.7 \pm 0.3) \times 10^{-19} {\rm
W\,m}^{-2}$ and ${\rm f}_{\rm [OIII]\,5007} = (15.1 \pm 0.6) \times
10^{-19} {\rm W\,m}^{-2}$.

The broad--band flux densities\footnote{Broad--band flux densities are
obtained by assuming the emission is flat across the filter, and are
placed at the central filter wavelength. Errors in this approximation are
comparable to or smaller than the other errors in measuring the flux
densities.} from the two HST images are marked on each spectrum. These
were obtained by convolving the HST images to a resolution comparable to
the seeing of the spectroscopic data, and then extracting the flux
densities through apertures of the same size and shape as the slit. For
3C34, the broad--band HST flux densities were corrected for line
contamination, using line fluxes measured from the spectral data (taking
into account the misplacement of the slit --- see below), thus providing
an estimate of the continuum flux density alone. It can be seen that, for
both spectra, these flux densities are almost a factor of two greater than
the values predicted from the spectrum, although the colour of the
spectrum matches the broad--band HST colours. Since this is true of the
spectra of both sources, the most likely explanation is that the slit was
misplaced from the centre of the targets by about 1.0 arcsec.

The dashed lines on Figure~\ref{specs} represent the spectral energy
distributions derived in Section~\ref{undergal} from stellar synthesis
fits to the broad--band flux densities, and normalised to match the
observed spectra. The spectral energy distribution derived for 3C34
provides a good match apart from the strong emission lines, which are
excited by photoionisation from the nucleus. Although the observed galaxy
`a' spectrum is dominated by noise, it is consistent with the overall
shape of the fitted SED, including the 4000\ang\ break.

\section{An emission knot on the radio axis?}
\label{jetclint}

\subsection{Evidence from the radio properties}
\label{radprops}

The radio image of 3C34 (Figure~\ref{hstims}b) indicates the possible
presence of radio jets in both lobes of the radio source. Garrington
\etal\ \shortcite{gar91} and Johnson \etal\ \shortcite{joh95} have
interpreted the two knots within the eastern lobe as being associated with
the radio jet; they have flatter spectral indices than the surrounding
lobe material, and the magnetic field is aligned along the direction to
the core, characteristic of radio jets in FRII sources. Directly opposite
these, the western arm shows an extension from the core and two slightly
enhanced regions of radio emission along a line towards the more southerly
of the two radio hotspots (hotspot `s'). Many pieces of evidence indicate
that hotspot `s' is the current primary hotspot in the western lobe: (i)
the present axis of the radio jets, as defined by the radio knots and the
western nuclear radio jet, is oriented towards this hotspot; (ii) the
8.4~GHz A--array observations (Figure~\ref{hstims}a) show that it is the
more compact of the two western hotspots; (iii) the 1.4 -- 4.8~GHz spectral
indices of hotspots `s' and `n' are $0.83\pm0.03$ and $0.92\pm0.03$
respectively, the flatter spectral index of hotspot `s' indicating that it
is younger; for comparison the spectral index of the eastern `hotspot' is
$0.82\pm0.03$.

The jet therefore appears to be currently pointing towards hotspot `s',
and not passing through object `a'. Given this, we must consider the
nature of hotspot `n'. Is it the site of an old impact of a precessing jet
\cite{sch82}, which at one time passed along the line through object `a',
or is it a `splatter--spot' where material flowing out of the primary
hotspot `s' strikes the opposite side of the radio cocoon \cite{wil85}?
The former hypothesis is supported by evidence for precession of the jet
in the eastern lobe of the source: the A--array data
(Figure~\ref{hstims}a) show that the emission from the end of the eastern
lobe is confused, with evidence for at least two hotspots; in addition,
the current jet direction in the eastern lobe, as defined by the two knots
of emission, points directly along the line from the active hotspot `s' in
the western lobe through the radio core, but does not point towards the
eastern hotspots, consistent with precession of the jet. 

Cox \etal\ \shortcite{cox91} showed that if a jet is steadily precessing
it initially makes a glancing impact upon the cocoon wall, causing it to
curve round but continue to feed the same hotspot. As it precesses
further, the jet eventually strikes the cocoon wall at a sharp enough
angle to generate a new primary hotspot, upstream of the old primary. This
is consistent with the relative locations of hotspots `s' and `n'. It
could also explain the `arc' of radio emission to the south of hotspot `s'
in the A--array image; the jet may have precessed slightly beyond this
hot--spot but is still continuing to feed it. Note also that the angle of
precession required for the jet to move from pointing towards hotspot `n'
to pointing towards hotspot `s' ($\lta 10^{\circ}$) is comparable to the
precession angles required to account for the observed asymmetries in
powerful double radio sources \cite{bes95a}. It therefore seems likely
that, at some previous time, the jet was pointing towards the hotspot `n'.

Despite the unavailability of a redshift, there is strong circumstantial
evidence that object `a' is associated with the radio source, rather than
merely being a foreground or background object along the line of
sight. The deep radio map (Figure~\ref{hstims}b) shows an enhanced region
of radio emission lying to the north of object `a'. The radio spectral
index of this region is not as steep as that of the rest of the radio
lobe, and increases away from the hotspot, indicating a region of rapid
backflow from the hotspot \cite{blu94}. This backflow loops around to the
north of object `a' rather than passing though it, consistent with object
`a' lying within the radio lobe, and the relativistic electrons flowing
out from the hotspot avoiding this region of higher gas density.

The western lobe of this source has a higher Faraday depolarisation than
the eastern lobe, and high resolution depolarisation mapping by Johnson
\etal\ \shortcite{joh95} has shown this to be associated with a
`depolarisation silhouette' which lies directly at the position of object
`a'. The 21cm to 6cm depolarisation measure ($DP^{21}_6$) is defined as
the ratio of the percentage polarisation at 21cm to that at 6cm. In a 5
arcsec long region, with a pear--shaped morphology almost identical to
that of object `a', the depolarisation measure has a value of $DP^{21}_6
\lta 0.1$ to 0.2, as compared with that of the surrounding lobe of
$DP^{21}_6 \gta 0.5$. Johnson \etal\ \shortcite{joh95} associate this
depolarisation with a cluster galaxy lying in front of the lobe, but it
would also be consistent with object `a' lying within the radio lobe.

\subsection{Optical and infrared properties of the knot}
\label{optprops}

\begin{figure}
\centerline{
\psfig{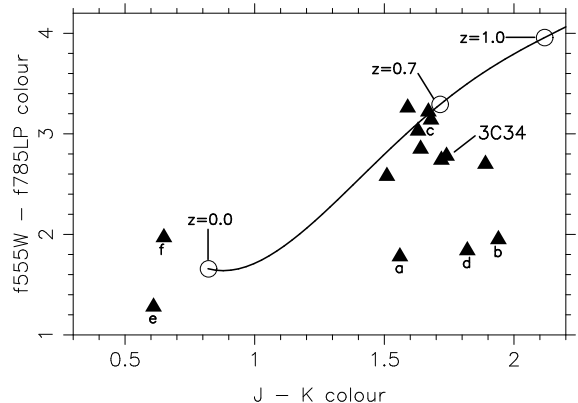}}
\caption{\label{optircols} A plot of optical vs infrared colour for
galaxies with K $< 19$ which lie within 250~kpc of the radio galaxy
3C34. Galaxies are plotted as a filled triangle, with typical errors of
$\pm 0.2$ in each coordinate. The solid line represents the colour track
obtained by redshifting a standard, current--epoch giant elliptical
galaxy. The open circular markers along this line indicate redshifts of
0.0, 0.7 and 1.0.}
\end{figure}

In Figure~\ref{optircols} we plot the f555W$-$f785LP
colour\footnote{Again, in the case of 3C34 which possesses strong line
emission, the f555W and f785LP flux densities have been corrected for line
contamination.}  against the J$-$K colour for all the galaxies within
250~kpc of 3C34 that have a K magnitude brighter than K = 19, that is,
those bright enough to have their magnitudes measured to an accuracy $\lta
0.2$ magnitudes. In this analysis, all magnitudes have been measured
through a 4 arcsec diameter aperture, and corrected for galactic
extinction using the extinction maps of Burstein and Heiles
\shortcite{bur82a}. The three galaxies, `a', `b' and `c', labelled on
Figure~\ref{hstims}a, are cross--referenced on Figure~\ref{optircols}; the
galaxies labelled `d', `e' and `f' on Figure~\ref{optircols} lie outside the
area shown in Figure~\ref{hstims}.

The near--infrared emission from these galaxies is dominated by their old
stellar population and, out to a redshift $z \sim 1.5$, the infrared J$-$K
colour is a fairly strong function of the redshift of the galaxy. For
passively evolving galaxies, the optical colour is also determined by the
redshift of the galaxy, but any active flat--spectrum components
(eg. star--formation, scattering etc) that may exist within the galaxies
will make a large contribution to the optical and ultraviolet
emission. Therefore, for galaxies of a given infrared colour, the optical
f555W$-$f785LP colour can be used as an indicator of any activity within
them. On Figure~\ref{optircols} we display the colour track obtained by
redshifting the spectrum of a standard elliptical galaxy, (produced using
the stellar synthesis codes provided by Bruzual and Charlot 1993,1997),
and assuming that there is no evolution of the stellar
population.\nocite{bru93,bru97}

It can be seen that the majority of the galaxies lie towards the upper
right corner of this plot, close to the expected locus of a standard
elliptical galaxy at the same redshift as 3C34. This would be consistent
with them being members of a cluster surrounding 3C34 \cite{mcc88}, and
shows that, at best, they are only passively evolving. A group of three
galaxies, `a', `b' and `d', possess the same infrared J$-$K colours as the
main group, suggesting that they may also be cluster galaxies. Their
optical colours are, however, a magnitude bluer and so they must be
optically active. Object `a' is marginally the bluest of this set of
galaxies. The difference in colour of object `a' from the spectrum of a
standard galaxy is apparent from a comparison of the images in
Figures~\ref{azoom} and~\ref{34zoom}. The second bluer galaxy, `b' (see
Figure~\ref{hstims}a), lies close to the boundary of the radio lobe, and
is elongated parallel to this transverse bow--shock, which may have some
bearing on its bluer colour. The only other galaxy with K $<19$ that lies
projected within the radio lobe is the bright galaxy `c' to the north of
3C34 (see Figure~\ref{hstims}a); this lies on the edge of the radio tail
emission, so would not be expected to be brightened. It is interesting to
compare the optical activity seen in galaxies `a' and `b' with the
observation of R\"ottgering \etal\ \shortcite{rot96a} that, to account for
the excess of companions along the direction of the radio axis in a sample
of ultra--steep spectrum radio sources, galaxies along the radio axis
would have to be brightened in the optical waveband by up to 2 magnitudes.

The third blue galaxy, `d', is over 200~kpc from 3C34, well away from the
radio axis, and fairly symmetrical. Some other mechanism must be
responsible for its colour. The two galaxies `e' and `f' which lie towards
the lower left corner of Figure~\ref{optircols} are likely to be
foreground galaxies; they are 2 to 3 magnitudes brighter in the f555W
waveband than the other galaxies. 
\smallskip

Based upon the optical, infrared and radio evidence, we conclude that it
is likely that, underlying the active flat spectrum emission from object
`a', lies a galaxy at the same redshift as 3C34.

\subsection{The galaxy underlying object `a'}
\label{undergal}

It is reasonable to assume that object `a' was an ordinary galaxy in the
cluster containing 3C34 before the radio source induced the flat spectrum
active component in some way. The infrared K--image, which is relatively
unaffected by the enhanced ultraviolet emission, shows the passively
evolving old stellar population of the galaxy, whilst the f555W filter HST
image is dominated by the optically bright regions corresponding to the
induced active emission.  Comparison of the locations of the optical and
infrared emission indicates that the centre of the underlying galaxy lies
in the gap between the two highly elongated emission regions.

\begin{figure}
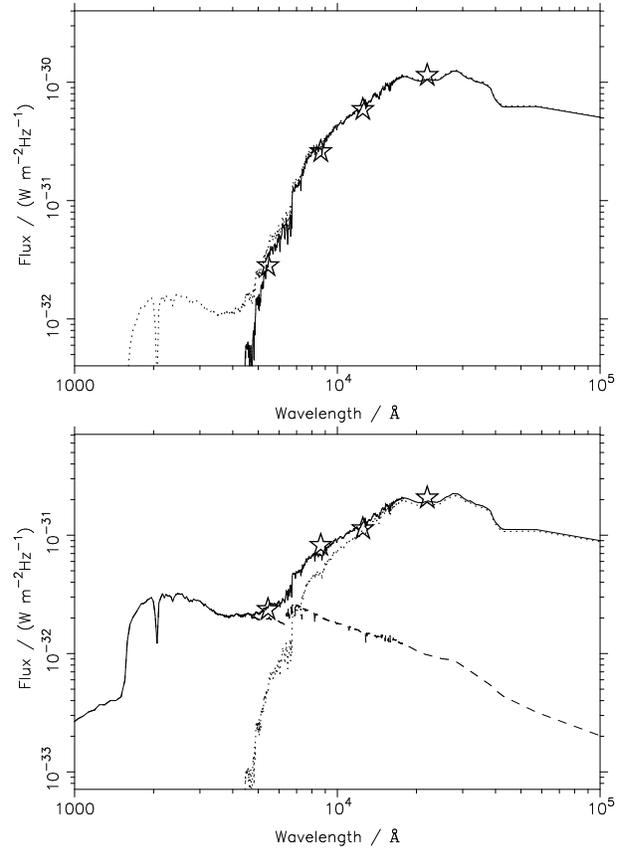

\centerline{
\psfig{figure=3c34fig6a.ps,clip=,angle=-90,width=8cm}
}
\centerline{
\psfig{figure=3c34fig6b.ps,clip=,angle=-90,width=8cm}
}
\caption{\label{sedfits} (a -- top) Solid line -- a fit to the SED of the
galaxy associated with 3C34, for a 5.5~Gyr old stellar population. Dotted
line -- addition of a 20~Myr old starburst (see Section~\ref{concs}). (b
-- bottom) Dotted line --- a fit to the underlying old stellar population
of galaxy `a' using a 5.5~Gyr old stellar population. It can be seen that
this falls short at ultraviolet wavelengths, requiring addition of a flat
spectrum component. Dashed line --- the SED of a 1~Myr starburst observed
5~Myr after it ended (see Section~\ref{stars}). Solid line --- a fit to
the total SED of galaxy `a' using the sum of the old stellar population
and the starburst. The stars indicate the broad--band HST flux
densities. These are corrected in (a) for strong line contamination. See
text for more details.}
\end{figure}

We can use the K--magnitude to normalise a fit of an old galaxy spectrum
to the spectral energy distribution (SED) of galaxy `a' at infrared
wavelengths, using the stellar synthesis codes of Bruzual and Charlot
(1993,1997).  \nocite{bru93,bru97} We adopt a Scalo \shortcite{sca86}
initial mass function with an upper mass cut--off at 65\,$M_{\odot}$. To
estimate the age of the galaxy, we first modelled the central cluster
radio galaxy. The colours and images of this galaxy (Figure~\ref{34zoom})
suggest that it is a passively evolving giant elliptical galaxy, and so we
can fit the SED using a single population. The old stellar population is
modelled as having an exponentially decreasing star--formation rate with
an e--folding time of 0.25\,Gyr. It can be seen in Figure~\ref{sedfits}a
(solid line) that such a stellar population at an age of $\sim 5.5$\,Gyr
(corresponding to a formation redshift of $z_{\rm f} \approx 10$) provides
a good fit to the broad--band HST flux densities.

To model galaxy `a', we assume that the underlying population has the same
age as that of the central radio galaxy, that is, 5.5~Gyr: a good match to
the infrared flux densities is provided using a galaxy mass of $1.65
\times 10^{10} M_{\odot}$ (see Figure~\ref{sedfits}b, dotted line). This
fit falls below the detected flux density at ultraviolet wavelengths, and
so an active flat--spectrum component is required to account for the blue
emission. This flat spectrum component contributes about 80\% of the flux
density detected in the f555W image. The total flux density measured using
this filter through a 4 arcsec diameter aperture centred on `a' is $(3.3
\pm 0.6) \times 10^{-32}$\,W\,Hz$^{-1}$m$^{-2}$.  Interestingly, the flux
density within apertures tightly surrounding the three bright regions of
emission in object `a' amounts to $(2.75 \pm 0.4) \times
10^{-32}$\,W\,Hz$^{-1}$m$^{-2}$, or about 85\% of the total. This is
comparable to the predicted value, especially since a significant fraction
of the light of the old galaxy will underlie these active regions and will
therefore have been included. The conclusion that can be drawn from this
is that these three bright components correspond to the active flat
spectrum emission regions.

A comparison of the flux density expected from the old stellar population
in the infrared J band with that actually observed indicates that the flat
spectrum component makes a small ($\sim 15\%$) contribution to the J--band
flux density. This may account for the slightly disturbed nature of this
image. Surprisingly, the same procedure indicates that the active emission
should only contribute about 30\% of the light through the f785LP HST
filter, with the old stellar population being responsible for the
remaining 70\%; the HST image appears to be more dominated by extended
aligned emission regions than by a passively evolving galaxy. At first
sight this would seem to suggest that, instead of being a passive
elliptical galaxy, the underlying galaxy itself must be somewhat extended
along the radio axis. However, it must be remembered that the active
emission regions have a much higher surface brightness than the more
diffuse underlying galaxy, and also that a fraction of the flux density
from a symmetrical galaxy would underlie the bright active emission
regions.

To test whether this observation is consistent with the underlying galaxy
being symmetrical, we attempted to remove the active component of the
emission in the f785LP image, in order to see what remains. To achieve
this, we used the f555W image as a template of where the active emission
lies, and scaled it until the flux density of the flat spectrum
contribution at this wavelength matched the predicted flux density of that
component in the f785LP observation. This scaled image was then subtracted
from the f785LP image to remove the 30\% active emission in this filter;
note that due to the presence of some galaxian component in the f555W
image, this also involved unavoidable subtraction of some of the
underlying galaxy light. The subtracted image was quite noisy, and so was
convolved with a 0.3 arcsec gaussian; this only smoothes the resultant
image slightly, and has little effect upon either the image resolution or
the qualitative nature of the resultant image.

\begin{figure}
\centerline{
\psfig{figure=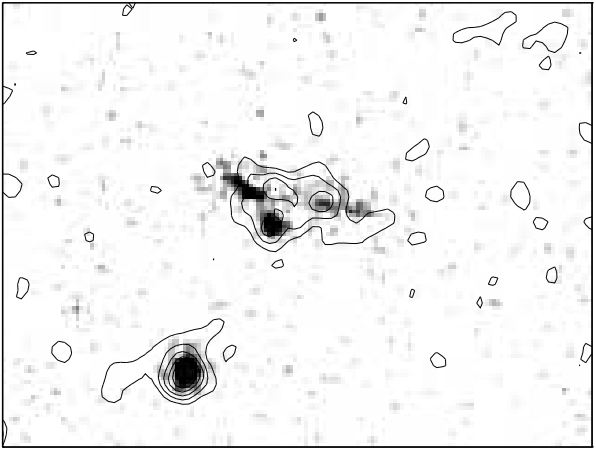,clip=,angle=90,width=8cm}
}
\caption{\label{3c34oldgal} An image of the `old stellar population'
component of the f785LP image. This is created by scaling the f555W image
(which is dominated by the active flat spectrum emission), and subtracting
it from the f785LP image to remove the predicted 30\% active flux from
that image. The contours show the remaining flux, from the underlying
galaxy. Relative to the first contour level, the contour levels are
1,2,3,4,5,6. This image is not inconsistent with a smooth symmetrical
galaxy underlying the active emission regions. The f555W image is plotted
in greyscale.}
\end{figure}

This smoothed image is displayed in Figure~\ref{3c34oldgal} as a contour
plot overlying the f555W image. Although the resultant image does not
follow a completely smooth galaxian profile, displaying three peaks of
emission, these peaks are present only at low significance. It is also
clear that this emission does not follow the extended blue emission seen
in the greyscale. The centre of this image lies directly over the centre
of the K--band image and if this image is convolved to the seeing of the
infrared observations then its appearance matches exactly that of the K
image. 

It is not possible to discount the possibility that what we are actually
observing here is a collision between two galaxies within the 3C34
cluster. In this scenario, the various peaks within the galaxy in
Figure~\ref{3c34oldgal} would correspond to the central regions of the
colliding galaxies, whilst the extended blue emission would be due to
tails of gas resulting from the tidal interactions of the merging
galaxies. However, the position of this merger directly along the axis of
the radio source, coupled with the alignment of the tidal tails along
the radio jet direction makes this possibility highly unlikely.

More likely is that an ordinary galaxy does underlie object `a', and that
active emission has been induced in this in some way by the radio source.
In the following sections we assume this to be the case, and discuss
possible mechanisms for producing this aligned emission.

\section{Alignment mechanisms}
\label{aligns}

\subsection{Scattering of quasar light by electrons or dust}
\label{scatter}

According to some unification schemes of radio galaxies and quasars
\cite{bar89}, these two populations of radio sources may arise from the
same parent population viewed at different orientations. In this model,
quasars have their radio axis orientated within about 45 degrees of the
line of sight, enabling us to observe their active galactic nuclei, whilst
radio galaxies have their axis orientated towards the plane of the sky,
and a torus of material obscures their central emission regions. In radio
galaxies, scattering of the obscured quasar light by dust or electrons
will produce polarised optical emission, which has been observed in many
sources (eg. Cimatti \etal\ \shortcite{cim96} and references
therein). This scattering will occur not only along the radio jet
direction, but from the whole cone within which the quasar light is
emitted. Even if enhanced by the presence of dust or electrons associated
with galaxy `a', the morphology of the scattered light should track the
underlying galaxy, rather than producing a linear feature.

If we assume that all of the flux from the `active' regions of object `a'
is associated with scattered light, we can obtain a lower limit to
the flux incident on this region from the quasar. We employ a similar
method to that used by other authors (eg. Eales and Rawlings 1990; van
Breugel and Dey 1993)\nocite{eal90,bre93}. 

The scattered luminosity, $L_{\rm scat}$, detected from object `a' is
related to the incident quasar luminosity, $L_{\rm Q}$, by:

\begin{equation}
\label{scateqn}
L_{\rm scat} \sim L_{\rm Q} \frac{R^2 {\rm sin}^2\theta }{D_{\rm proj}^2}
f_{\rm c} \left (1 - e^{-\tau}\right ) G(\pi -\theta)
\end{equation}

\noindent where $L_{\rm scat}$ ($\sim 7.1 \times
10^{19}$\,W\,Hz$^{-1}$sr$^{-1}$) and $L_{\rm Q}$ are measured in
W\,Hz$^{-1}$sr$^{-1}$; $R$ ($\approx$ 15~kpc) is the characteristic size
of galaxy `a'; $D_{\rm proj}$ ($\approx$ 120~kpc) is its projected
distance from 3C34; $\theta$ is the angle between the optical cone axis of
the quasar emission and the line of sight; $f_{\rm c}$ is the covering
factor of material within galaxy `a', given by $f_{\rm c} \approx f_{\rm
v}^{2/3}$, where $f_{\rm v}$ is the volume filling factor of the material;
$\tau$ is the optical depth for scattering through galaxy `a'; and
$G(\theta)$ is the differential scattering cross--section of the
scattering material. We adopt a value of $\theta = 90^{\circ}$, meaning
that the radio axis lies in the plane of the sky. This minimises the
amount of scattering required and is consistent with the symmetry of
3C34\footnote{Many pieces of evidence suggest that 3C34 lies more or less
in the plane of the sky: (i) Jet candidates are identified in both
lobes. (ii) The two lobes are roughly equal in length and, with the
exception of the depolarisation silhouette, also in Faraday
depolarisation. (iii) Both lobes show well--defined inner edges with a
clear gap between the lobes corresponding to the position of the cluster
core --- if the source was at a significant angle to the plane of the sky,
this gap would have been blurred.}.

In the case of scattering by free electrons, $G(\theta) =
\frac{3}{16\pi}\left (1 + {\rm cos}^2\theta\right )$, and the optical
depth to scattering through galaxy `a' is given by $\tau \sim n_{\rm e}
\sigma_{\rm T} R f_{\rm v}^{1/3}$, where $n_{\rm e}$ is the number density
of electrons and $\sigma_{\rm T} = 6.65 \times 10^{-29}$\,m$^2$ is the
Thompson scattering cross--section. The number density of electrons is
obtained by averaging the mass of gas throughout the volume of the clouds
within galaxy `a'; $n_{\rm e} \sim M_{\rm gas}/(R^3 f_{\rm v} m_{\rm p})$,
where $m_{\rm p}$ is the atomic mass.

Assuming that $L_{\rm Q} \sim 10^{23}$\,W\,Hz$^{-1}$sr$^{-1}$
\cite{ost89}, and noting that since $\tau$ is small, $(1 - e^{-\tau})
\approx \tau$, we can substitute these values into equation~\ref{scateqn}
and derive the mass of gas required to produce the observed scattering
luminosity: $M_{\rm gas} \sim 2 \times 10^{11} M_{\odot}$. This is
much higher than estimates of the mass of warm ($T \sim 10^{4}$K)
emission line gas within galaxies, which are in the range $10^7$ to $10^9
M_{\odot}$ (McCarthy 1993 and references therein), thus ruling out warm
electrons as a possible scattering agent. Estimates of the hot X--ray gas
masses in nearby galaxies lie in the range $10^8$ to $10^{11} M_{\odot}$
\cite{fab89b}, with the upper end of the range corresponding to the most
massive galaxies. The K--band magnitude of galaxy `a' indicates that it is
a fairly average galaxy (eg. compare Figures~\ref{azoom}d and
and~\ref{34zoom}d), and so it is extremely unlikely that it contains over
$10^{11} M_{\odot}$ of gas (cf. Section~\ref{undergal} where we derived a
mass of $1.65 \times 10^{10} M_{\odot}$ of old stars in the
galaxy). Scattering by hot electrons is therefore unlikely to play an
important role in this object.

Dust scattering is more efficient than electron scattering in the
ultraviolet waveband, since the scattering cross--section of dust
particles is of the same order as the geometric cross--section. Following
the same procedure as for electron scattering, and adopting values of
$G(90^{\circ}) \sim 0.05$ for the differential scattering cross--section,
$a_{\rm d} \sim 10^{-7}$\,m for the radius of the dust grains, and
$\rho_{\rm d} \sim 3000$\,kg\,m$^{-3}$ for their density \cite{whi92}, we
derive an estimate for the dust mass required in galaxy `a' of $M_{\rm
dust} \gta 2.5 \times 10^7 M_{\odot}$. For a galactic gas--to--dust ratio
of about 150 \cite{whi92} this give a total mass of gas in the warm--phase
(ie. excluding the X--ray gas) of $M_{\rm gas} \sim 4 \times 10^9
M_{\odot}$. This, again, is a high mass of warm--phase gas compared to
typical measured values, and compared to the mass of stars derived in
Section~\ref{undergal}.  In addition, quasar light incident upon such a
high mass of warm ionised gas would be expected to produce emission lines
of sufficient strength to be observed in the spectrum
(Figure~\ref{specs}a).  The absence of such lines therefore means that
dust--scattering is also unlikely to be of importance in object `a'.

\subsection{Inverse Compton scattering / Optical synchrotron emission}
\label{compt}

Inverse Compton scattering and optical synchrotron emission should appear
brightest at the peaks in the radio emission, rather than in a single
elongated feature at a radio minimum. These mechanisms have been found to
be important in some local sources, such as M87 \cite{bir91}, but are not
generally found to be important in the powerful radio galaxies (see also
van Breugel and Dey \shortcite{bre93} for 3C285).

The flux density due to optical synchrotron emission can be calculated
directly from the radio flux density by assuming a constant spectral index
$\alpha$:

\begin{displaymath}
\frac{f_{\rm sync}(\nu)}{f_{\rm r}(\nu _{\rm r})} \sim \left
(\frac{\nu_{\rm r}}{\nu} \right ) ^{\alpha}
\end{displaymath}

\noindent which, in the region of object `a', gives $f_{\rm sync} \sim 1.4
\times 10^{-34}$\,W\,Hz$^{-1}$m$^{-2}$. This will contribute less than 1\%
of the flux density observed from this region. In practice the continuum
spectrum is expected to steepen as the break frequency is passed
\cite{hug91}, leading to an even lower limit to the level of optical
synchrotron flux.

The contribution to the optical flux density from up--scattered microwave
background photons can be calculated using the equations derived by Daly
(1992a,b) \nocite{dal92a,dal92b}:

\begin{displaymath}
\frac{f_{\rm ic}(\nu)}{f_{\rm r}(\nu _{\rm r})} \sim 1.6 \times 10^{-12}
\frac{(1 + z)^{(1 - \alpha)}}{\epsilon^{(1+\alpha )}} k \left
[\frac{7.5 \times 10^{17}}{\nu}\frac{\nu _{\rm r}}{178\,{\rm MHz}} \right
]^{\alpha}
\end{displaymath}
 
\noindent where $f_{\rm ic}(\nu)$ and $f_{\rm r}(\nu _{\rm r})$ are the
flux densities at the optical frequency $\nu$ and the radio frequency
$\nu_{\rm r}$ respectively; $\epsilon ^2 = 0.092 B_\perp^2 / (1 + z)^4$ is
a parameterisation of the magnetic field in the lobe (measured in $\mu
G$); $\alpha$ is the radio spectral index, $f_{\rm r}(\nu ) \propto \nu
^{-\alpha}$; and $k$ is a constant which depends upon $\alpha$, having a
value of 160 for $\alpha = 1$.

For the radio emission close to object `a', $f_{\rm r} \sim 1.5$\,mJy at
$\nu _{\rm r} = 8.4$\,GHz, $B \approx 10\,\mu G$, and $\alpha \approx 1$
\cite{joh95}.  Therefore through the f555W filter, $f_{\rm ic} \sim 2
\times 10^{-34}$\,W\,Hz$^{-1}$m$^{-2}$, which is again much smaller than
the observed flux density from object `a'.

\subsection{Nebular continuum emission}
\label{nebular}

Dickson \etal\ \shortcite{dic95} recently suggested that a significant
percentage of the ultraviolet flux from high redshift radio galaxies may
be associated with nebular continuum emission (free--free, free--bound and
bound--bound interactions). If gas within galaxy `a' were collisionally
excited by shocks associated with the radio jet, these emission mechanisms
would result in a morphology similar to that observed. The strength of nebular
continuum emission is, however, strongly correlated with line strengths,
and the failure to detect any emission lines indicates that this mechanism
is unlikely to play a significant role in galaxy `a'.

\subsection{Young Stars}
\label{stars}

Star formation induced by the passage of the radio jet through galaxy `a'
could give rise to a morphology similar to that observed. There is,
however, a problem with the star formation hypothesis in that the flux of
ionising photons from the most massive stars in a newly formed stellar
population should give rise to significant line emission. This problem can
be circumvented by considering the ageing of a starburst population. In
Section~\ref{radprops} we discussed the fact that the jet currently points
at hotspot `s', and is no longer passing through galaxy `a'. It is
therefore reasonable to suppose that star formation may no longer be
continuing in this galaxy. Indeed, if the period of star formation was a
temporary phenomenon occurring only whilst the active hotspot region of
the radio source was advancing through the galaxy (see Best \etal\ (1996b)
for a discussion of this model), then the age of the starburst can be
estimated using radio spectral ageing arguments.

The radio spectrum is interpreted in terms of an ageing population of
electrons, with the higher spectral index in the material further from the
hotspots being due to the lower synchrotron break frequency in the older
electron population. Assuming (i) a constant magnetic flux density, (ii) a
standard power--law injection spectrum for the electrons, and (iii) that
the electrons are isotropised on time--scales much shorter than their
radiative lifetime, then the evolution of the break frequency with time is
given by the equation \cite{pac70,liu92}:

\begin{displaymath}
(\nu_{\rm T} / {\rm GHz}) = 2.5 \times 10^3 \left [\frac{(B / {\rm
nT})^{1/2}}{(B / {\rm nT})^2 + (B_{\rm MWB} / {\rm nT})^2}\right ]^2 (t /
{\rm Myr})^{-2}
\end{displaymath}
 
\noindent where $\nu_{\rm T}$ is the break frequency, below which the spectrum
is a power--law, and above which it steepens; $B$ is the magnetic field
strength; $B_{\rm MWB} = 0.315(1+z)^2$ nT is the equivalent magnetic strength
of the cosmic microwave background radiation; and $t$ is the spectral age
of the electrons, that is, the time that has elapsed since they were last
accelerated.

Blundell \shortcite{blu94} has estimated the strength of the magnetic
field in 3C34 using equipartition arguments (eg. Alexander and Leahy
1987)\nocite{ale87}, and derived values of 0.5 to 1\,nT in the inner
regions of the radio lobes, that is, the regions closest to the host radio
galaxy. These values are consistent with the estimates of a cluster
magnetic field strength of 0.4\,nT needed to produce the observed Faraday
depolarisation, if the depolarisation is due to hot cluster gas of typical
cluster density \cite{joh95}. Blundell also fitted spectra to different
regions of the source, and calculated the break frequencies at different
locations. In the inner regions of the radio lobes these were typically 5
to 10\,GHz.

Inserting these values into the above equation, the spectral ages of the
oldest electrons, which are assumed to lie in the inner regions of the
radio lobes furthest from the hotspots, are $9 \times 10^6$ to $3 \times
10^7$ yr, giving hotspot advance speeds of order $0.05c$. The distance
between galaxy `a' and the hotspot indicates that the starburst is
observed about $6 \times 10^6$ years after it was excited by the passage of the
radio jet through the galaxy. This corresponds roughly to the main
sequence lifetime of a 20\,$M_{\odot}$ star, and so all stars of greater
mass than this will have completed their evolution by the observed epoch.

Although infrequent in number, the strong dependence of stellar luminosity
on mass means that these massive stars play a significant role in the
starburst luminosity, especially at ultraviolet wavelengths. Returning to
Figure~\ref{sedfits}b, we attempt to fit the excess ultraviolet emission
required to account for the measured flux densities by adding a starburst
population, observed $6 \times 10^6$ years after it occurred to the
underlying galaxy SED. It is seen that a mass of $1.0 \times 10^8
M_{\odot}$ of young stars provides a good match to all of the observed
data points. This corresponds to a star formation rate of about
100\,$M_{\odot}$ per year during the 1\,Myr burst.

\smallskip

\begin{figure}
\centerline{
\psfig{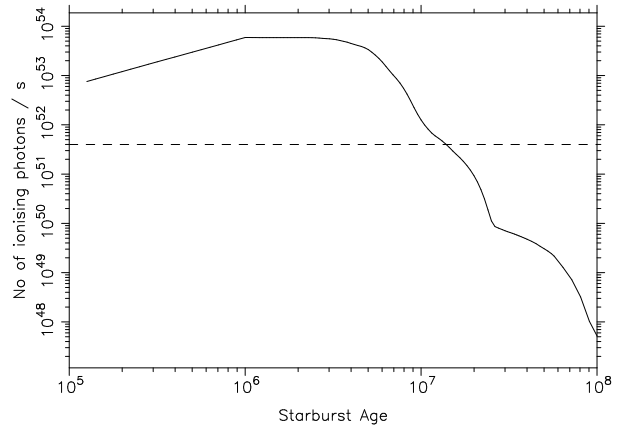}
}
\caption{\label{ionphots} A plot of the number of ionising photons against
age of the starburst (solid line). The dashed line shows the constant
level of ionising photons arising from the AGN.}
\end{figure}

In addition to dominating the starburst colour, the most massive stars,
being the hottest, emit the vast majority of the photons with energies
sufficient to ionise hydrogen and oxygen in these sources: the ionisation
potential of oxygen is 13.5\,eV, nearly identical to that of hydrogen.  In
Figure~\ref{ionphots} we show how the number of photons emitted per second
with sufficient energy to ionise hydrogen and oxygen, decreases rapidly
with age for the starburst considered above in galaxy `a'. For comparison
we also plot an estimate of the number of ionising photons that would be
intercepted by galaxy `a' from an obscured quasar nucleus in 3C34,
assuming it to have properties typical of 3CR quasars at that redshift
(ie. $\sim 10^{54}$ ionising photons per second \cite{mcc93} emitted
within a cone of opening half--angle 45$^{\circ}$).  It can be seen that,
at the distance of galaxy `a' from the nucleus, the majority of its
ionising photons initially arises from the starburst, but the number of
these decreases rapidly falling by nearly a factor of 100 in $10^7$
years.

It is possible to estimate the emission line flux that would be observed
from galaxy `a' if it absorbs all of the ionising photons from the active
nucleus. Using the simplest assumption, each ionising photon will
eventually produce one \la\ photon, and so the luminosity of \la\ due to
the central AGN is about $9 \times 10^{-20}$\,W\,m$^{-2}$. Using a \la\ to
[OII]~3727 ratio of 5, taken from the mean of a large number of radio
galaxy spectra in which the emission lines are also excited predominantly
by photoionisation from the AGN \cite{mcc88}, this would mean that the
[OII] line flux should be observed at about $1.8 \times
10^{-20}$\,W\,m$^{-2}$, which is between one and two times the noise
level on the spectrum of galaxy `a' (Figure~\ref{specs}a).

In practice, not all of the ionising photons will be absorbed, due to two
factors: firstly, the covering fraction of the emission line gas is likely
to be less than one; secondly, the absorption rate is limited by the
availability of neutral hydrogen atoms, which depends upon the
recombination rate of the H$^+$ ions and electrons (eg. Osterbrock
1989).\nocite{ost89} If the emission line gas is evenly distributed,
producing a covering factor of unity, then the low hydrogen density
dictates a slow recombination rate and limits the number of ionisations to
$\sim 2 \times 10^{49}$s$^{-1}$. For emission line clouds with properties
characteristic of those around high redshift radio sources, that is $n_e
\sim 10^8$m$^{-3}$ and a total mass of warm gas of $\sim 10^8 M_{\odot}$
\cite{hec91}, the volume filling factor is only $\sim 10^{-5}$ meaning
that the vast majority of the ionising photons will pass through the
galaxy unabsorbed. The larger number of ionising photons associated with a
newly formed starburst would produce line fluxes just detectable on the
spectrum in Figure~\ref{specs}a. However, the decrease in ionising photons
that occurs by the age of $\sim 6 \times 10^6$ years at which this
starburst is observed, produces a corresponding decrease in line fluxes,
and reduces the expected line flux to below the noise level.

\section{Discussion}
\label{concs}

The results of Section~\ref{stars} show that the lack of line emission
from object `a' is entirely consistent with an ageing starburst. Indeed,
the rapid decrease of line emission with starburst age leads us to ask a
different question: rather than ask why object `a' does not show line
emission, we should instead ask why Minkowski's object and the region 09.6
in the lobe of 3C285 show so much, if they were formed by an interaction
in a similar way to object `a'. In the case of Minkowski's object, the
object lies at the position where the radio jet disrupts, and is therefore
likely to be the result of a relatively recent interaction. The knot 09.6 in
the lobe of 3C285, however, lies at a similar distance behind the hotspot
as object `a'. Van Breugel and Dey \shortcite{bre93} noted that it `has an
emission line spectrum typical of a starburst galaxy', but then suggested
that it was an instantaneous starburst with an age of $70 \pm 30$ Myr,
based on the spectral shape and the size of the 4000\AA\
break. Figure~\ref{ionphots} shows that if this were the case, few high
ionisation emission lines would be expected in the spectrum.

A reasonable fit to the spectrum of 09.6 can be obtained assuming a
younger ($\lta 10~$Myr) starburst involving about 10\% of the material
from an older ($\sim 1~$Gyr) galaxy. In this case, the size of the
4000\AA\ break is provided mainly by the older stars. This starburst age
would also be more consistent with the radio spectral ageing of this
source \cite{ale87}, but even in this case it would be expected to have
line emission as weak as that in galaxy `a'.

In the case of 3C34, the powerful jets were capable of driving through
galaxy `a', thereby inducing a rapid, short--duration, burst of star
formation with a highly elongated morphology tracking the passage of the
jet. By contrast, the radio jet associated with 3C285, which is roughly a
factor of 100 lower in power than that in 3C34, does not seem to penetrate
the knot 09.6: the knot is fairly symmetrical apart from being
edge--brightened, particularly in line emission, on the upstream side; the
continuum is also bluest on the side facing the nucleus; the radio jet
bends at a bright radio knot close to the point of impact with 09.6,
indicating that the jet may have been disrupted and deflected by the knot,
rather than passing through it. The interaction of the weaker jet in this
source appears to have induced star formation only on the side of the knot
that it originally struck, resulting in a significantly different physical
situation as compared with that of galaxy `a'. It is plausible that this
weaker interaction may result in a less extreme but more prolonged
starburst, with low levels of on--going star formation. Perhaps the bend
in the radio jet at the knot near 09.6 results in momentum flux continuing
to be incident upon the cloud, which may be responsible for its continued
excitation. 

It is interesting to compare our results with a recent observation by
Cimatti \etal\ \shortcite{cim96} that a companion close to the radio
galaxy 3C324 ($z = 1.206$), and positioned along the radio axis, may
plausibly be undergoing (or have very recently undergone) a burst of
star--formation at a rate of $70 M_{\odot}$ yr$^{-1}$. The similarity of
this feature of 3C324 to the situation in 3C34 is striking.
 
The star formation rates suggested in these two cases can be compared to
the values required to account for the aligned blue structures in the host
galaxies of 3CR radio sources. Lilly and Longair \shortcite{lil84a}
accounted for the blue excess of high redshift radio sources using star
formation rates of `several' solar masses per year for a duration of
$10^7$ to $10^8$ years, whilst Dunlop \etal\ \shortcite{dun89} suggest
that about 1\% of the mass of the galaxy would need to be involved in the
starburst. Each of these models predicts a comparable mass of young stars
to the model presented here.  Rees \shortcite{ree89} and Begelman and
Cioffe \shortcite{beg89} have considered the star formation rates induced
by radio bow shocks expanding through a two--phase medium with reasonable
filling factor and star--formation efficiency, and have derived star
formation rates of order $100 M_{\odot}$\,yr$^{-1}$.

\bigskip

Our preferred interpretation for the recent history of 3C34 is as follows:

\begin{enumerate}
 
\item The radio jets were recently pointing towards hotspot `n' in the
western lobe, and the southernmost hotspot in the eastern lobe. The
backflow of relativistic electrons from hotspot `n' passed to the north of
galaxy `a'.

\item The radio jet passed through the halo of galaxy `a' in the western
lobe, and supersonic shocks associated with its passage induced a massive
burst of star formation. The most massive stars died out within the few
$\times 10^6$ years between the onset of this starburst and the epoch at
which we observe it, with the consequences that the colour of the
starburst is redder than expected for a currently active star--forming
region, and there are few ionising photons to produce line emission.

\item Precession of the jets has given rise to a disconnection event in the
western lobe, with the currently active hotspot `s' closer to the
nucleus. The relativistic electrons continue to backflow through the
evacuated region to the north of galaxy `a'. In the eastern lobe,
precession has resulted in the formation of new hotspots. The two radio
knots lie along the current jet axis.

\item Shocks associated with the transverse expansion of the radio cocoon
may have induced some star formation in other galaxies close to the radio
axis, accounting for their blue optical colour. 
\end{enumerate}

In this model, the interaction of the radio jet with galaxy `a' induces
about 0.6\% of the mass of the galaxy to be converted into young stars in
a time span of 1\,Myr. To investigate the practicality of such a model, we
can compare these predictions with the observations of a nearby system,
the Cartwheel galaxy. In the Cartwheel galaxy, the passage of a companion
through the spiral galaxy at small impact parameter has resulted in a
rapidly propagating ring--shaped structure within the spiral disk (eg. see
Higdon 1995\nocite{hig95}). Violent star formation is seen to be occurring
at the shock front within this ring, and Kennicutt \shortcite{ken83}
derived a star formation rate of $67 M_{\odot}$ yr$^{-1}$. This ring has
been expanding for 300\,Myr, and so a large fraction of the mass of this
galaxy must have been involved in the starburst.

The passage of the companion through the Cartwheel galaxy is, in many
ways, similar to the passage of the radio jet through galaxy `a'. In
galaxy `a', the associated bow shocks will be more powerful, and so will
pass through the galaxy more quickly leading to the much shorter duration
of the starburst. Despite the lower mass of this galaxy as compared to the
Cartwheel, it is not unreasonable to expect that similar rates of star
formation may occur. 

Perhaps an even better comparison is that of nearby `E + A'
galaxies. These are galaxies which are dominated by a young stellar
component, frequently involving up to 10\% of the galaxy mass, but which,
like galaxy `a', lack the emission lines characteristic of on--going star
formation (eg. Zabludoff \etal\ \shortcite{zab96} and references
therein). It is thought that `E + A' galaxies may be the result of violent
galaxy interactions; the interaction of the radio jet with galaxy `a' will
produce a qualitatively similar result. Interestingly, `E + A' galaxies
have distinctive spectra dominated by strong Balmer absorption lines. Our
spectroscopic data have neither the spectral resolution nor the
signal--to--noise necessary to detect these lines, but better
signal--to--noise spectra at higher spectral resolution may provide direct
proof of the existence, or otherwise, of an ageing starburst.

The separation of galaxy `a' from the active galactic nucleus in this
source suggests that the properties of this galaxy may not be typical of
the aligned regions of the host radio galaxies. It possesses less emission
line, scattered light and nebular continuum contributions, as compared
with the extended emission regions in the radio galaxies. If, however, the
radio jets are powerful enough to induce star formation over 100~kpc from
the central engine, then the star formation process must also be of great
importance in producing the aligned structures within the host galaxies.

One question that may be asked is why there is no evidence for star
formation occuring within the host galaxy of 3C34?. The answer fits in
remarkably well with our results from the study of the eight galaxies in
the sample which lie within the redshift range $1 \lta z \lta 1.3$
\cite{bes96a}. The optical morphologies of these eight galaxies are seen
to evolve strongly with radio size: small radio sources are composed of
many bright knots of emission tightly aligned along the radio axis, whilst
those with more extended radio emission contain only one or two bright
components and generally have less extended optical emission.

Best \etal\ \shortcite{bes96a} showed that this morphological evolution
would be consistent with jet induced star formation models whereby cold
clumps of gas are induced to collapse and form stars as the radio
components pass through the host galaxy, and these stars then evolve
passively. The relatively short lifetimes of the most massive luminous
stars, coupled with relaxation of the star forming regions within the
gravitational potential of the host galaxy leads to a significant decrease
in the starburst luminosity over the lifetime of the radio source.

3C34 is one of the largest radio sources in our sample and any star
formation that occurred within the host galaxy will be about $2 \times
10^7$ yr old, meaning that the starburst will be well beyond its peak
luminosity. By this age it is dominated, even in the rest--frame near
ultraviolet, by the much more massive old stellar population: addition of
a $2 \times 10^8 M_{\odot}$ starburst at this age, has only a small effect
on the infrared and optical spectral energy distribution (see
Figure~\ref{sedfits}a). If present, it may be responsible for the slight
east--west extension seen at low significance in the f555W image
(Figure~\ref{34zoom}a), but our observations cannot distinguish whether or
not such a population is present in this galaxy. Observations at 3000\AA ,
however, would easily distinguish whether an ageing starburst exists in
older radio galaxies.

By contrast, star formation induced by the passage of the jet through
galaxy `a' is relatively recent: the structures bear a striking
similarity to those seen within many of the smaller (younger) 3C radio
galaxies. In addition, the significantly lower gravitational potential of
galaxy `a' will allow the strikingly aligned morphology of its young stars
to remain visible for a longer period.
 
\section*{Acknowledgements} 
\label{acknowl}

This work is based on observations with the NASA/ESA Hubble Space
Telescope, obtained at the Space Telescope Science Institute, which is
operated by AURA Inc., under contract from NASA.  The National Radio
Astronomy Observatory is operated by AURA Inc., under co-operative
agreement with the National Science Foundation. The authors wish to thank
Richard Saunders and Malcolm Bremer for kindly taking the spectrum of
galaxy `a', and Paddy Leahy for providing us with his radio map of 3C34.
We thank the referee for helpful comments. PNB acknowledges support from
PPARC. HJAR acknowledges support from an EU twinning project, a programme
subsidy granted by the Netherlands Organisation for Scientific Research
(NWO) and a NATO research grant.

\label{lastpage}
\bibliography{pnb} 

\begin{thebibliography}{}

\bibitem[\protect\citename{Alexander \& Leahy{\ }}{1987}]{ale87}
Alexander~P.,  Leahy~J.~P.,  1987, MNRAS, 225, 1

\bibitem[\protect\citename{Barthel{\ }}{1989}]{bar89}
Barthel~P.~D.,  1989, ApJ, 336, 606

\bibitem[\protect\citename{Begelman \& Cioffi{\ }}{1989}]{beg89}
Begelman~M.~C.,  Cioffi~D.~F.,  1989, ApJ, 345, L21

\bibitem[\protect\citename{Best et~al.{\ }}{1995}]{bes95a}
Best~P.~N.,  Bailer~D.~M.,  Longair~M.~S.,    Riley~J.~M.,  1995, MNRAS, 275,
  1171

\bibitem[\protect\citename{Best et~al.{\ }}{1996}]{bes96a}
Best~P.~N.,  Longair~M.~S.,    R{\"o}ttgering~H. J.~A.,  1996, MNRAS, 280, L9

\bibitem[\protect\citename{Best et~al.{\ }}{1997}]{bes97c}
Best~P.~N.,  Longair~M.~S.,    R{\"o}ttgering~H. J.~A.,  1997, MNRAS:
  submitted.

\bibitem[\protect\citename{Biretta et~al.{\ }}{1991}]{bir91}
Biretta~J.~A.,  Stern~C.~P.,    Harris~D.~E.,  1991, AJ, 101, 1632

\bibitem[\protect\citename{Blundell{\ }}{1994}]{blu94}
Blundell~K.~M.,  1994, Ph.D. thesis, University of Cambridge

\bibitem[\protect\citename{Brodie et~al.{\ }}{1985}]{bro85}
Brodie~J.,  Bowyer~S.,    McCarthy~P.~J.,  1985, ApJ, 293, L59

\bibitem[\protect\citename{Bruzual \& Charlot{\ }}{1993}]{bru93}
Bruzual~G.,  Charlot~S.,  1993, ApJ, 405, 538

\bibitem[\protect\citename{Bruzual \& Charlot{\ }}{1997}]{bru97}
Bruzual~G.,  Charlot~S.,  1997

\bibitem[\protect\citename{Burstein \& Heiles{\ }}{1982}]{bur82a}
Burstein~D.,  Heiles~C.,  1982, AJ, 87, 1165

\bibitem[\protect\citename{Chambers et~al.{\ }}{1987}]{cha87}
Chambers~K.~C.,  Miley~G.~K.,    {van Breugel}~W. J.~M.,  1987, Nat, 329, 604

\bibitem[\protect\citename{Cimatti et~al.{\ }}{1996}]{cim96}
Cimatti~A.,  Dey~A.,  {van Breugel}~W.,  Antonucci~R.,    Spinrad~H.,  1996,
  ApJ, 465, 145

\bibitem[\protect\citename{Cox et~al.{\ }}{1991}]{cox91}
Cox~C.~I.,  Gull~S.~F.,    Scheuer~P. A.~G.,  1991, MNRAS, 252, 558

\bibitem[\protect\citename{Daly{\ }}{1990}]{dal90}
Daly~R.~A.,  1990, ApJ, 355, 416

\bibitem[\protect\citename{Daly{\ }}{1992a}]{dal92b}
Daly~R.~A.,  1992a, ApJ, 399, 426

\bibitem[\protect\citename{Daly{\ }}{1992b}]{dal92a}
Daly~R.~A.,  1992b, ApJ, 386, L9

\bibitem[\protect\citename{{De Young}{\ }}{1989}]{dey89}
{De Young}~D.~S.,  1989, ApJ, 342, L59

\bibitem[\protect\citename{Dey \& Spinrad{\ }}{1996}]{dey96}
Dey~A.,  Spinrad~H.,  1996, ApJ, 459, 133

\bibitem[\protect\citename{Dickson et~al.{\ }}{1995}]{dic95}
Dickson~R.,  Tadhunter~C.,  Shaw~M.,  Clark~N.,    Morganti~R.,  1995, MNRAS,
  273, L29

\bibitem[\protect\citename{Dunlop et~al.{\ }}{1989}]{dun89}
Dunlop~J.~S.,  Guiderdoni~B.,  Rocca-Volmerange~B.,  Peacock~J.,    Longair~M.,
   1989, MNRAS, 240, 257

\bibitem[\protect\citename{Eales \& Rawlings{\ }}{1990}]{eal90}
Eales~S.~A.,  Rawlings~S.,  1990, MNRAS, 243, 1P

\bibitem[\protect\citename{Fabbiano{\ }}{1989}]{fab89b}
Fabbiano~G.,  1989, ARA\&A, 27, 87

\bibitem[\protect\citename{Fanaroff \& Riley{\ }}{1974}]{fan74}
Fanaroff~B.~L.,  Riley~J.~M.,  1974, MNRAS, 167, 31P

\bibitem[\protect\citename{Garrington et~al.{\ }}{1991}]{gar91}
Garrington~S.~T.,  Conway~R.~G.,    Leahy~J.~P.,  1991, MNRAS, 250, 171

\bibitem[\protect\citename{Heckman et~al.{\ }}{1991}]{hec91}
Heckman~T.~M.,  Lehnert~M.~D.,  {van Breugel}~W. J.~M.,    Miley~G.~K.,  1991,
  ApJ, 370, 78

\bibitem[\protect\citename{Higdon{\ }}{1995}]{hig95}
Higdon~J.~L.,  1995, ApJ, 455, 524

\bibitem[\protect\citename{Hughes{\ }}{1991}]{hug91}
Hughes~P.~A.,  1991, Beams and Jets in Astrophysics.
Cambridge Astrophysics Series 19, Cambridge University Press

\bibitem[\protect\citename{Johnson et~al.{\ }}{1995}]{joh95}
Johnson~R.~A.,  Leahy~J.~P.,    Garrington~S.~T.,  1995, MNRAS, 273, 877

\bibitem[\protect\citename{Kennicutt{\ }}{1983}]{ken83}
Kennicutt~R.~C.,  1983, ApJ, 272, 54

\bibitem[\protect\citename{Laing et~al.{\ }}{1983}]{lai83}
Laing~R.~A.,  Riley~J.~M.,    Longair~M.~S.,  1983, MNRAS, 204, 151

\bibitem[\protect\citename{Lauer{\ }}{1989}]{lau89}
Lauer~T.~R.,  1989, PASP, 101, 445

\bibitem[\protect\citename{Lilly \& Longair{\ }}{1984}]{lil84a}
Lilly~S.~J.,  Longair~M.~S.,  1984, MNRAS, 211, 833

\bibitem[\protect\citename{Liu et~al.{\ }}{1992}]{liu92}
Liu~R.,  Pooley~G.,    Riley~J.~M.,  1992, MNRAS, 257, 545

\bibitem[\protect\citename{McCarthy{\ }}{1988}]{mcc88}
McCarthy~P.~J.,  1988, Ph.D. thesis, University of California, Berkeley

\bibitem[\protect\citename{McCarthy{\ }}{1993}]{mcc93}
McCarthy~P.~J.,  1993, ARA\&A, 31, 639

\bibitem[\protect\citename{McCarthy et~al.{\ }}{1987}]{mcc87}
McCarthy~P.~J.,  {van Breugel}~W. J.~M.,  Spinrad~H.,    Djorgovski~S.,  1987,
  ApJ, 321, L29

\bibitem[\protect\citename{Osterbrock{\ }}{1989}]{ost89}
Osterbrock~D.~E.,  1989, Astrophysics of Gaseous Nebulae and Active Galactic
  Nuclei.
Mill Valley CA: University Science Books

\bibitem[\protect\citename{Pacholczyk{\ }}{1970}]{pac70}
Pacholczyk~A.~G.,  1970, Radio Astrophysics.
Freeman, San Francisco

\bibitem[\protect\citename{Perley{\ }}{1989}]{per89}
Perley~R.~A.,  1989, in Perley~R.~A.,  Schwab~F.~R.,   Bridle~A.~H.,  eds, ASP
  Conf. Ser. Vol 6, Synthesis Imaging in Radio Astronomy.
Astronomical Society of the Pacific, p.~259

\bibitem[\protect\citename{Rees{\ }}{1989}]{ree89}
Rees~M.~J.,  1989, MNRAS, 239, 1P

\bibitem[\protect\citename{R{\"o}ttgering et~al.{\ }}{1996}]{rot96a}
R{\"o}ttgering~H. J.~A.,  West~M.~J.,  Miley~G.~K.,    Chambers~K.~C.,  1996,
  A\&A, 307, 376

\bibitem[\protect\citename{Scalo{\ }}{1986}]{sca86}
Scalo~J.~M.,  1986, Fund. Cosmic Phys., 11, 1

\bibitem[\protect\citename{Scheuer{\ }}{1982}]{sch82}
Scheuer~P. A.~G.,  1982, in Heeschen~D.~S.,  Wade~C.~M.,  eds, IAU Symp 97:
  Extragalactic Radio Sources.
Reidel, Dordrecht, p.~163

\bibitem[\protect\citename{{van Breugel} \& Dey{\ }}{1993}]{bre93}
{van Breugel}~W. J.~M.,  Dey~A.,  1993, ApJ, 414, 563

\bibitem[\protect\citename{{van Breugel} et~al.{\ }}{1985}]{bre85}
{van Breugel}~W. J.~M.,  Filippenko~A.~V.,  Heckman~T.~M.,    Miley~G.~K.,
  1985, ApJ, 293, 83

\bibitem[\protect\citename{Whittet{\ }}{1992}]{whi92}
Whittet~D. C.~B.,  1992, Dust in the galactic environment.
Institute for Physics Publishing, Bristol

\bibitem[\protect\citename{Williams \& Gull{\ }}{1985}]{wil85}
Williams~A.~G.,  Gull~S.~F.,  1985, Nat, 313, 34

\bibitem[\protect\citename{Zabludoff et~al.{\ }}{1996}]{zab96}
Zabludoff~A.~I.,  Zaritsky~D.,  Lin~H.,  Tucker~D.,  Hashimoto~Y.,
  Shectman~S.~A.,  Oemler~A.,    Kirshner~R.~P.,  1996, ApJ, 466, 104

\end{thebibliography}
\bibliographystyle{mn} 

\end{document}